\documentclass[aps,prl,twocolumn,showpacs,preprintnumbers,amsmath,amssymb,refmerge,superscriptaddress]{revtex4}
\usepackage{graphicx}
\usepackage{dcolumn}
\usepackage{longtable}
\newcommand{\ra}{\rightarrow}

 \newcommand{\lam}{\Lambda}

\begin{document}


\title{Observation of $B^+\ra\bar{\Xi}_c^0\Lambda_c^+$ and Evidence for 
$B^0\ra\bar{\Xi}_c^-\Lambda_c^+$}

\affiliation{Budker Institute of Nuclear Physics, Novosibirsk}
\affiliation{Chonnam National University, Kwangju}
\affiliation{University of Cincinnati, Cincinnati, Ohio 45221}
\affiliation{Gyeongsang National University, Chinju}
\affiliation{University of Hawaii, Honolulu, Hawaii 96822}
\affiliation{High Energy Accelerator Research Organization (KEK), Tsukuba}
\affiliation{Hiroshima Institute of Technology, Hiroshima}
\affiliation{Institute of High Energy Physics, Chinese Academy of Sciences, Beijing}
\affiliation{Institute of High Energy Physics, Vienna}
\affiliation{Institute for Theoretical and Experimental Physics, Moscow}
\affiliation{J. Stefan Institute, Ljubljana}
\affiliation{Kanagawa University, Yokohama}
\affiliation{Korea University, Seoul}
\affiliation{Kyungpook National University, Taegu}
\affiliation{Swiss Federal Institute of Technology of Lausanne, EPFL, Lausanne}
\affiliation{University of Ljubljana, Ljubljana}
\affiliation{University of Maribor, Maribor}
\affiliation{University of Melbourne, Victoria}
\affiliation{Nagoya University, Nagoya}
\affiliation{Nara Women's University, Nara}
\affiliation{National Central University, Chung-li}
\affiliation{National United University, Miao Li}
\affiliation{Department of Physics, National Taiwan University, Taipei}
\affiliation{H. Niewodniczanski Institute of Nuclear Physics, Krakow}
\affiliation{Nippon Dental University, Niigata}
\affiliation{Niigata University, Niigata}
\affiliation{Nova Gorica Polytechnic, Nova Gorica}
\affiliation{Osaka City University, Osaka}
\affiliation{Osaka University, Osaka}
\affiliation{Peking University, Beijing}
\affiliation{University of Pittsburgh, Pittsburgh, Pennsylvania 15260}
\affiliation{Princeton University, Princeton, New Jersey 08544}
\affiliation{University of Science and Technology of China, Hefei}
\affiliation{Shinshu University, Nagano}
\affiliation{Sungkyunkwan University, Suwon}
\affiliation{University of Sydney, Sydney NSW}
\affiliation{Tata Institute of Fundamental Research, Bombay}
\affiliation{Toho University, Funabashi}
\affiliation{Tohoku Gakuin University, Tagajo}
\affiliation{Tohoku University, Sendai}
\affiliation{Department of Physics, University of Tokyo, Tokyo}
\affiliation{Tokyo Institute of Technology, Tokyo}
\affiliation{Tokyo Metropolitan University, Tokyo}
\affiliation{University of Tsukuba, Tsukuba}
\affiliation{Virginia Polytechnic Institute and State University, Blacksburg, Virginia 24061}
\affiliation{Yonsei University, Seoul}
 \author{R.~Chistov}\affiliation{Institute for Theoretical and Experimental Physics, Moscow} 
   \author{K.~Abe}\affiliation{High Energy Accelerator Research Organization (KEK), Tsukuba} 
   \author{I.~Adachi}\affiliation{High Energy Accelerator Research Organization (KEK), Tsukuba} 
   \author{H.~Aihara}\affiliation{Department of Physics, University of Tokyo, Tokyo} 
   \author{Y.~Asano}\affiliation{University of Tsukuba, Tsukuba} 
   \author{V.~Aulchenko}\affiliation{Budker Institute of Nuclear Physics, Novosibirsk} 
 \author{T.~Aushev}\affiliation{Institute for Theoretical and Experimental Physics, Moscow} 
   \author{S.~Bahinipati}\affiliation{University of Cincinnati, Cincinnati, Ohio 45221} 
   \author{A.~M.~Bakich}\affiliation{University of Sydney, Sydney NSW} 
   \author{V.~Balagura}\affiliation{Institute for Theoretical and Experimental Physics, Moscow} 
% ...
\collaboration{The Belle Collaboration}

 
\begin{abstract}

We report the first observation of the decay $B^+\ra\bar{\Xi}_c^0\Lambda_c^+$ with a significance of $8.7~\sigma$ and evidence for the decay 
$B^0\ra\bar{\Xi}_c^-\Lambda_c^+$ with a significance of $3.8~\sigma$.
The product
${\cal B}(B^+\ra\bar{\Xi}_c^0\Lambda_c^+)\times{\cal B}(\bar{\Xi}_c^0\ra\bar{\Xi}^+\pi^-)$
is measured to be $(4.8^{+1.0}_{-0.9}\pm 1.1\pm 1.2)\times 10^{-5}$,  
and
${\cal B}({B^0}\ra\bar{\Xi}_c^-\Lambda_c^+)\times{\cal B}(\bar{\Xi}_c^-\ra\bar{\Xi}^+\pi^-\pi^-)$ is measured to be
$(9.3^{+3.7}_{-2.8}\pm 1.9\pm 2.4)\times 10^{-5}$. 
The errors are statistical, systematic and the error of the $\Lambda_c^+\ra pK^-\pi^+$
branching fraction, respectively.
The decay $B^+\ra\bar{\Xi}_c^0\Lambda_c^+$ is the first example of
a two-body exclusive $B^+$ decay into two charmed baryons.
The data used for this analysis was accumulated at the $\Upsilon(4S)$ resonance,
using the Belle detector at the $e^+ e^-$ asymmetric-energy collider KEKB.
The integrated luminosity of the data sample is equal to $357\,~\mathrm{fb}^{-1}$,
corresponding to $386\times 10^{6}$ $B{\bar B}$ pairs.

\end{abstract}


\maketitle

{\renewcommand{\thefootnote}{\fnsymbol{footnote}}}
\setcounter{footnote}{0}

A number of $B$-meson decay modes to final states containing  baryons have been
observed, including $b\ra c\bar{u} d$ decays with either one final-state charmed meson
(e.g. $B^0\ra \bar{D^0} p\bar{p}$~\cite{belle_dpp})
or a charmed baryon
(e.g. $B^+\ra\bar{\Lambda}_c^- p\pi^+$~\cite{belle_kichimi_1}), and charmless baryonic
decays~\cite{belle_ppk} that proceed via $b\ra s$ or $b\ra u$ transitions.
Two-body baryonic decay modes are found to have lower branching fractions than
multi-body modes and, in the latter,
near-threshold enhancements are observed in the baryon-pair invariant mass spectra~\cite{kichimi_review}.
Some theoretical models attribute these phenomena to baryonic form factors
that are large for multi-body modes~\cite{theory_b_to_baryons}.

\begin{figure}[h]
\centering
\includegraphics[width=0.45\textwidth]{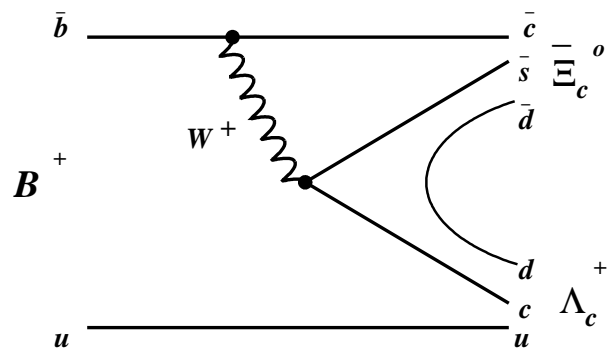}
\caption{ 
The quark diagram for the $B^+\ra\bar{\Xi}_c^0\Lambda_c^+$ decay.}
\label{diagramm}
\end{figure}

Recently, Belle reported examples of baryonic decays that proceed via $b\ra c{\bar c}s$ transitions:
$B^-\ra J/\psi\Lambda\bar{p}$~\cite{jpsilamp} and $B\ra\Lambda_c^+\bar{\Lambda}_c^- K$~\cite{kolya}.
To date, however, nothing is experimentally known about two-body exclusive $B$ decays
to two charmed baryons, which would also proceed through $b\ra c{\bar c}s$ transitions.
An example of such a decay is
 $B^+\ra\bar{\Xi}_c^0\Lambda_c^+$, which would proceed via the
quark-diagram shown in Fig.~\ref{diagramm}.
This two-body $B$ decay mode,
like $B\ra\Lambda_c^+\bar{\Lambda}_c^- K$,
would produce a ``wrong-sign'' ${\Lambda}_c^+$, in contrast to all other known
$B$ decay modes that only have $\bar{\Lambda}_c^-$'s in the final state~\cite{pdg}.
Recently the BaBar collaboration has measured the inclusive yield of
(wrong-sign) ${\Lambda}_c^+$'s from $B$ decays~\cite{babar_incl_ws_lamc}.
It was suggested that this type of $B$ decay might be a substantial component
of baryonic $b\ra c{\bar c}s$ transitions and could have an important influence
on
the determination of the charm particle yield per $B$ decay~\cite{dunietz}.

For exclusive two-body baryonic modes, a theoretical model based on QCD sum rules predicts
${\cal B}(B\ra\bar{\Xi}_c\Lambda_c^+)\sim 10^{-3}$\cite{chernyak}.
Experimental measurements of $B\ra\bar{\Xi}_c\Lambda_c^+$
test theoretical predictions and provide additional information on the
dynamics of two-body baryonic $B$ decays.

In this Letter we report the first observation of 
$B^+\ra\bar{\Xi}_c^0\Lambda_c^+$ and evidence for 
$B^0\ra\bar{\Xi}_c^-\Lambda_c^+$ decays.
Charge conjugation is implied here and throughout the paper.
The analysis is performed using data collected
with the Belle detector at the KEKB asymmetric-energy $e^+e^-$
collider~\cite{kekb}.  The data sample consists of $357$~fb$^{-1}$ collected at the
$\Upsilon(4S)$ resonance, which corresponds to $386\times 10^6$ $B\bar B$ pairs.

The Belle detector is a large-solid-angle magnetic spectrometer that consists of a silicon
vertex detector (SVD), a 50-layer central drift chamber (CDC), an array of aerogel threshold
Cherenkov counters (ACC), a barrel-like arrangement of time-of-flight scintillation counters
(TOF), and an electromagnetic calorimeter (ECL) comprised of CsI(Tl) crystals located
inside a superconducting solenoid coil that provides a 1.5 T magnetic field. An iron flux-return located outside of the coil is instrumented to detect $K^0_L$ mesons and to identify muons
(KLM). The detector is described in detail elsewhere~\cite{belle_detector}. 
Two different inner detector configurations were used. 
For the first sample of 152 million $B\bar{B}$ pairs (Set I), a 2.0 cm
radius beampipe and a 3-layer silicon vertex detector were used; for the latter 234 million
$B\bar{B}$ pairs (Set II), a 1.5 cm radius beampipe, a 4-layer silicon detector and a small-cell inner
drift chamber were used~\cite{svd2}.
We use GEANT-based Monte Carlo (MC) simulation to model 
the response of the detector and determine the efficiency~\cite{montecarlo}.

We select charged pions, kaons and protons that originate from the region
$dr<1$~cm, $|dz|<4$~cm, where $dr$ and $dz$ are the distances of
closest approach to the interaction point in the plane
perpendicular to the beam axis ($r-\phi$ plane) and along the beam direction, respectively.
Pions, kaons and protons are identified 
using a likelihood ratio method, which combines information 
from the TOF system and ACC counters with $dE/dx$ measurements in the CDC~\cite{id_pkpi}.

In this analysis we reconstruct the following decay modes: $\Xi_c^0\ra\Xi^-\pi^+$ and $\Lambda K^-\pi^+$, $\Xi_c^+\ra\Xi^-\pi^+\pi^+$, $\Lambda_c^+\ra pK^-\pi^+$,
$\Xi^-\ra\Lambda\pi^-$ and $\lam \rightarrow p \pi^-$. 
For $\lam \rightarrow p \pi^-$, we fit the $p$ and $\pi$ tracks to a common vertex and  
require an invariant mass in a $\pm 5\,\mathrm{MeV}/c^2$
interval around the $\Lambda$ mass. 
The distance between the $\lam$ decay vertex position and interaction point (IP)
in the $r-\phi$ plane ($dr(\Lambda)$) is required to be greater than $0.05$~cm and 
the angle $\alpha_{\Lambda}$, between the $\lam$ momentum vector
and the vector pointing from the IP to the decay vertex,
must satisfy $\cos\alpha_{\Lambda} > 0.995$ for the case of $\Xi_c^0\ra\Lambda K^-\pi^+$. 
We make no requirements on $dr$ and $|dz|$ 
for tracks coming from $\Xi^-\ra\Lambda\pi^-$ and $\Lambda\ra p\pi^-$ decays.
For $\Xi^-\ra\Lambda\pi^-$, we fit the $\lam$ trajectory and the $\pi^-$ track to a common vertex and require a 
$\lam \pi^-$ invariant mass in a $\pm 5$~MeV/c$^2$
interval around the $\Xi^-$ mass. 
We require that the distance between the 
$\Xi^-$ decay vertex position and IP in the $r-\phi$ plane to be greater than $0.01$~cm. 
For the $\Lambda$'s coming from $\Xi^-$ in the decay $\Xi_c^0\ra\Xi^-\pi^+$ we apply the requirements, $dr(\Lambda)>0.5$~cm and $\cos\alpha_{\Lambda}>0.0$. 
For $\Lambda_c^+$, $\Xi_c^0$ and $\Xi_c^+$ we use mass windows that are $\pm 15\,\mathrm{MeV}/c^2$ around their nominal values. 
We use a large sample of inclusive $\Lambda$, $\Xi^-$, $\Xi_c^{+/0}$ and $\Lambda_c^+$ signals to verify that their mass peaks are well described by two Gaussians, corresponding to the core and tail of the distribution. 
The signal mass windows that are used in this analysis correspond to 
approximately $4\sigma$ for the core and $2\sigma$ for the tail Gaussian. 
The MC studies of the inclusive $\lam$, $\Xi^-$, $\Xi_c^{+/0}$ and $\Lambda_c^+$ 
signals show agreement with data.


\begin{table*}[t]
\begin{ruledtabular}
\caption {Summary of the fit results, efficiencies, products of branching fractions and statistical significances. 
For the $B^+$ for two $\bar{\Xi_c^0}$ modes 
the product of branching fractions is
${\cal B}(B^+\ra\bar{\Xi}_c^0\Lambda_c^+)\times{\cal B}(\bar{\Xi}_c^0\ra\bar{\Xi}^+\pi^-)$ since for $\bar{\Xi}_c^0\ra\bar{\Lambda} K^+\pi^-$ we use the ratio ${\cal B}(\Xi_c^0\ra\Lambda K^-\pi^+)/{\cal B}(\Xi_c^0\ra\Xi^-\pi^+)$ mentioned in the text. 
The uncertainties in the products of the branching ratios are statistical, systematic and the uncertainty of the $\Lambda_c^+\ra pK^-\pi^+$ branching fraction.}
\begin{tabular}{ccccc}
Decay Mode  & Yield & Efficiency($\%$) & Product of ${\cal B}$'s $(10^{-5})$  &
Significance \\
\hline
$B^+\ra\bar{\Xi}_c^0\Lambda_c^+$, $\bar{\Xi}_c^0\ra\bar{\Xi}^+\pi^-$ & $12.4^{+4.2}_{-3.3}$  & 1.14  & $5.6^{+1.9}_{-1.5} \pm 1.1\pm 1.5$  & $6.8\sigma$ \\
$B^+\ra\bar{\Xi}_c^0\Lambda_c^+$, $\bar{\Xi}_c^0\ra\bar{\Lambda} K^+\pi^-$ & $16.9^{+4.8}_{-4.0}$  & 2.04   & $4.0^{+1.1}_{-0.9}\pm 0.9\pm 1.0$  & $5.9\sigma$ \\
\hline
$B^+\ra\bar{\Xi}_c^0\Lambda_c^+$, simultaneous fit &  &  & $4.8^{+1.0}_{-0.9}\pm 1.1 \pm 1.2$   & $8.7\sigma$   \\
\hline
$B^0\ra\bar{\Xi}_c^-\Lambda_c^+$, $\bar{\Xi}_c^-\ra\bar{\Xi}^+\pi^-\pi^-$ & $8.3^{+3.3}_{-2.5}$  & 0.46 & $9.3^{+3.7}_{-2.8}\pm 1.9\pm 2.4$   &  $3.8\sigma$ \\
\end{tabular}
\label{table1}
\end{ruledtabular}
\end{table*}

The $B$ candidates ({\emph {i.e.} $\bar{\Xi}_c\Lambda_c^+$ combinations) 
are identified by their center of mass (c.m.) energy difference,
$\Delta E=\Sigma_i E_i - E_{\rm beam}$, and their beam-energy constrained mass,
$M_{\rm bc}=\sqrt{E^2_{\rm beam}-(\Sigma_i \vec{p}_i)^2}$, where $E_{\rm beam}=\sqrt{s}/2$
is the beam energy in the c.m. and $\vec{p}_i$ and $E_i$ are
the three-momenta and energies of the $B$ candidate's decay products.
We accept $B$ candidates with 
$M_{\rm bc}>5.2$~GeV$/c^2$ and $|\Delta E|<0.2$~GeV. 
To suppress the continuum background, we require the normalized 
Fox-Wolfram moment~\cite{fox2} $R_2$ to be less than 0.5.
We apply $|{\rm cos}\theta_B|<0.85$ for the $\Xi_c^0$ reconstruction in the 
$\Lambda K^-\pi^+$ mode, to suppress the combinatorial background. 
Here $\theta_B$ is the polar angle of the $B$-meson direction in the c.m.

The $\Delta E$ and $M_{\rm bc}$ distributions for the $B^+\ra\bar{\Xi}_c^0\Lambda_c^+$
candidates are shown in Figs.~\ref{f1} (a) and (b), 
where the two $\Xi_c^0$ modes are combined. 
We require $M_{\rm bc}>5.272$~GeV$/c^2$ ($|\Delta E|<0.025$~GeV) for the $\Delta E$ ($M_{\rm bc}$) projection~\cite{doublecounting}.
The hatched histograms in Figs.~\ref{f1} (a) and (b) 
show the sum of normalized 
$\Lambda_c^+$ and $\bar{\Xi}_c^0$ mass sidebands~\cite{sidebands_definition} 
 where no peaking structures are evident.
The superimposed curves are the results of a simultaneous two-dimensional binned 
maximum likelihood fit to the both $\Delta E$ versus $M_{\rm bc}$ 
distributions (for the two $\Xi_c^0$ channels) with a common value of 
${\cal B}(B^+\ra\bar{\Xi}_c^0\Lambda_c^+)\times{\cal B}(\bar{\Xi}_c^0\ra\bar{\Xi}^+\pi^-)\times{\cal B}(\Lambda_c^+\ra pK^-\pi^+)$. 
For this fit, we constrain the ratio ${\cal B}(\Xi_c^0\ra\Lambda K^-\pi^+)/{\cal B}(\Xi_c^0\ra\Xi^-\pi^+)$ to the recent Belle measurement of $1.07\pm 0.12\pm 0.07$~\cite{lesyak}.
To describe the signal we use Gaussians with 
means and widths fixed to the values obtained from MC.
The backgrounds in $\Delta E$ and $M_{\rm bc}$ are parametrized by a first-order polynomial 
and an ARGUS function~\cite{argus_function}, respectively.  
The fit gives a statistical significance of $8.7\sigma$ for the signal,  
where the statistical significance is defined as $\sqrt{-2{\rm ln}(L_0/L_{\rm max})}$, where $L_0$ and $L_{\rm max}$ are the likelihoods with the signal fixed at zero and at the fitted value, respectively.  
The region $\Delta E<-0.08$~GeV is excluded from the fit to avoid possible contributions from $B^{+/0}\ra\bar{\Xi}_c^0\Lambda_c^+\pi^{0/-}$ and $B^{0/+}\ra\bar{\Xi}_c^0\Sigma_c^{0/+}$, $\Sigma_c^{0/+}\ra\Lambda_c^+\pi^{-/0}$ decays, where the pion is undetected. 
The same fitting procedure applied separately for the two $\Xi_c^0$ modes gives 
$12.4^{+4.2}_{-3.3}$ (6.8~$\sigma$ significance) and $16.9^{+4.8}_{-4.0}$ (5.9~$\sigma$ significance) events 
for $B^+\ra\bar{\Xi}_c^0\Lambda_c^+$ followed by $\bar{\Xi}_c^0\ra\bar{\Xi}^+\pi^-$ and $B^+\ra\bar{\Xi}_c^0\Lambda_c^+$ followed by $\bar{\Xi}_c^0\ra\bar{\Lambda} K^+\pi^-$, respectively. 


As a cross-check of the $B^+\ra\bar{\Xi}_c^0\Lambda_c^+$ signal, we select events in
 the $B$-signal region of $|\Delta E|<0.025$~GeV and $M_{\rm bc}>5.272$~GeV/c$^2$ 
for two $\Xi_c^0$ modes and examine  the 
$\Lambda_c^+$ and $\bar{\Xi}_c^0$ mass distributions (Fig.~\ref{f1} (c) and (d)). 
For the $\Lambda_c^+$ ($\bar{\Xi}_c^0$) distribution we require $\bar{\Xi}_c^0$ ($\Lambda_c^+$) 
to be within $\pm 15\,\mathrm{MeV}/c^2$ of the nominal mass. We then fit each distribution 
with two Gaussians for the signal and a first-order polynomial to describe the background. 
The widths and means of the Gaussians are fixed to the values obtained from data as described above.
The fitted signal yields of $32.6\pm 7.2$ events for the $\Lambda_c^+$ and $29.4\pm 6.9$ events 
for the $\bar{\Xi}_c^0$ are in good agreement with the total signal yield for $B^+\ra\bar{\Xi}_c^0\Lambda_c^+$, including the two $\Xi_c^0$ decay modes. 

\begin{figure}[h]
\centering
\includegraphics[width=0.5\textwidth]{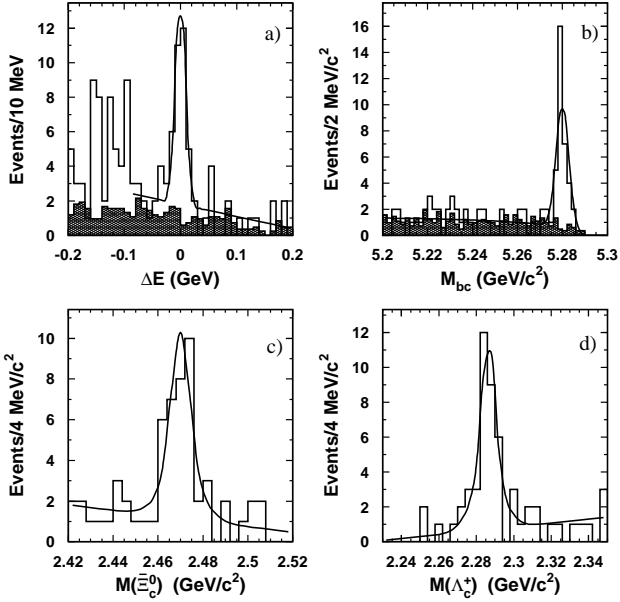}
\caption{ a) and b): The $\Delta E$(a)) and $M_{\rm bc}$(b)) distributions for 
the $B^+\ra\bar{\Xi}_c^0\Lambda_c^+$ candidates. The hatched histograms show the combined $\bar{\Xi}_c^0$ and $\Lambda_c^+$ mass sidebands normalized to the signal region. The excess around $\Delta E = -0.150$~GeV may be due to the contributions from $B^{+/0}\ra\bar{\Xi}_c^0\Lambda_c^+\pi^{0/-}$ and $B^{0/+}\ra\bar{\Xi}_c^0\Sigma_c^{0/+}$, $\Sigma_c^{0/+}\ra\Lambda_c^+\pi^{-/0}$ decays, where the pion is undetected. Therefore, we exclude this region from the fit. 
c) and d): The $\bar{\Xi}_c^0$ (c)) and $\Lambda_c^+$ (d)) mass distributions for the $B^+\ra\bar{\Xi}_c^0\Lambda_c^+$ candidates taken from the B-signal region of $|\Delta E|<0.025$~GeV and $M_{\rm bc}>5.272$~GeV/c$^2$. For the $\bar{\Xi}_c^0$ ($\Lambda_c^+$) distribution we require $\Lambda_c^+$ ($\bar{\Xi}_c^0$) to be within $\pm 15\,\mathrm{MeV}/c^2$ of the nominal mass. The overlaid curves are the fit results (see the text).
}
\label{f1} 
\end{figure}

\begin{figure}[h]
\centering
\includegraphics[width=0.5\textwidth]{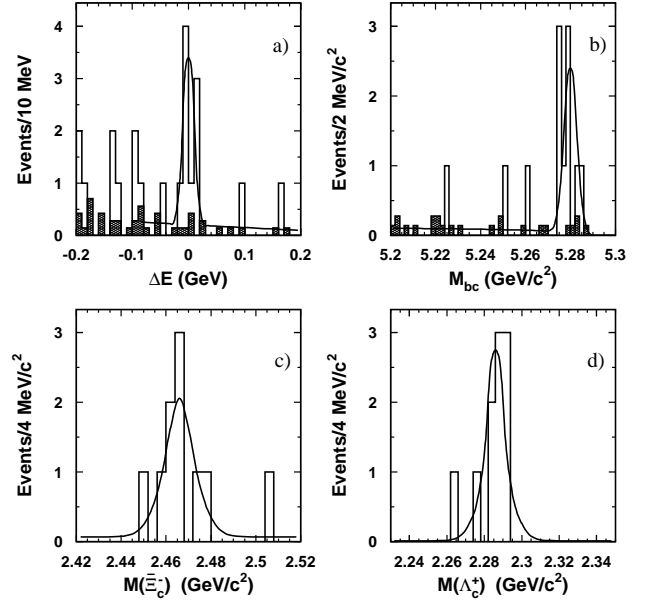}
\caption{ a) and b): The $\Delta E$(a)) and $M_{\rm bc}$(b)) distributions
for the $B^0\ra\bar{\Xi}_c^-\Lambda_c^+$ candidates. The hatched histograms show the combined $\bar{\Xi}_c^-$ and $\Lambda_c^+$ mass sidebands normalized to the signal region.  c) and d): The $\bar{\Xi}_c^-$ (c)) and $\Lambda_c^+$ (d)) mass distributions for the $B^0\ra\bar{\Xi}_c^-\Lambda_c^+$ candidates taken from the B-signal region of $|\Delta E|<0.025$~GeV and $M_{\rm bc}>5.272$~GeV/c$^2$. For the $\bar{\Xi}_c^-$ ($\Lambda_c^+$) distribution we require $\Lambda_c^+$ ($\bar{\Xi}_c^-$) to be within $\pm 15\,\mathrm{MeV}/c^2$ of the nominal mass. The overlaid curves are the fit results (see the text).
}
\label{f3} 
\end{figure}

The $B^0\ra\bar{\Xi}_c^-\Lambda_c^+$ mode is an isospin partner of 
the $B^+\ra\bar{\Xi}_c^0\Lambda_c^+$ mode. 
Therefore their branching fractions are expected to be of the same order of magnitude.
The $\Delta E$ and $M_{\rm bc}$ distributions for the $\bar{B^0}\ra\bar{\Xi}_c^-\Lambda_c^+$ 
candidates are shown in Figs.~\ref{f3} (a) and (b). 
The superimposed curves are the results of a two-dimensional binned maximum likelihood fit 
to the $\Delta E$ versus $M_{\rm bc}$ distribution. 
The fit gives $8.3^{+3.3}_{-2.5}$ signal events.
The signal significance is $3.8\sigma$, taking into account the systematic uncertainty from the signal and background parameterization.
The hatched histogram  shows the sum of the normalized $\Lambda_c^+$ and $\bar{\Xi}_c^-$ mass sidebands.
We apply the same procedure used for $B^+\ra\bar{\Xi}_c^0\Lambda_c^+$ 
to check the $\Lambda_c^+$ and $\bar{\Xi}_c^-$ signals as shown in Figs.~\ref{f3} (c) and (d). The fit gives $9.0\pm 3.0$ events for the $\Lambda_c^+$ and $8.4\pm 2.8$ events for the $\bar{\Xi}_c^-$. Both 
are in agreement with the ${B^0}\ra\bar{\Xi}_c^-\Lambda_c^+$ signal yield. 


Table~\ref{table1} summarizes the results of the fits for the 
$B^+$ and $B^0$ decays, the
reconstruction efficiencies including the ${\cal B}(\Lambda\ra p\pi^-)$,  
statistical significance of the signals and extracted 
products of branching fractions.
Here we use
${\cal B}(\Lambda_c^+\ra pK^-\pi^+)=(5.0\pm 1.3)\%$~\cite{pdg} and 
assume equal fractions of charged and neutral $B$ mesons produced in $\Upsilon(4S)$ decays.

The major sources of systematic error are
the uncertainties in the tracking
efficiency of $7\%$ ($1\%$ per track), $11\%$ in charged particle identification efficiency 
($1\%$ for pion, $2\%$ for kaon and $3\%$ for proton),
$5\%$ in $\Lambda$ finding,  $6\%$ in efficiency estimation due to MC statistics,
$10\%$ in the signal and background parameterization, and 
$13\%$ in ${\cal B}(\Xi_c^0\ra\Lambda K^-\pi^+)/{\cal B}(\Xi_c^0\ra\Xi^-\pi^+)$. 
Added in quadrature, these correspond to a total
systematic error of $23\%$ for $B^+\ra\bar{\Xi}_c^0\Lambda_c^+$ and $20\%$ for $\bar{B^0}\ra\bar{\Xi}_c^-\Lambda_c^+$.

In summary, we report the first observation of the $B^+\ra\bar{\Xi}_c^0\Lambda_c^+$ decay mode 
and the first evidence for the $B^0\ra\bar{\Xi}_c^-\Lambda_c^+$ decay mode.
The products of branching fractions
${\cal B}(B^+\ra\bar{\Xi}_c^0\Lambda_c^+)\times{\cal B}(\bar{\Xi}_c^0\ra\bar{\Xi}^+\pi^-)=(4.8^{+1.0}_{-0.9}\pm 1.1\pm 1.2)\times 10^{-5}$
and ${\cal B}({B^0}\ra\bar{\Xi}_c^-\Lambda_c^+)\times{\cal B}(\bar{\Xi}_c^-\ra\bar{\Xi}^+\pi^-\pi^-)=(9.3^{+3.7}_{-2.8}\pm 1.9\pm 2.4)\times 10^{-5}$ are measured with $8.7\sigma$ and $3.8\sigma$ significance, respectively.
These results and Belle's recent observation of $B\ra\Lambda_c^+\bar{\Lambda}_c^- K$~\cite{kolya} decays are the first examples of $B$ decays into two charmed baryons.
The branching fraction obtained for $B^+\ra\bar{\Xi}_c^0\Lambda_c^+$ together
with the theoretical predictions for ${\cal B}(\Xi_c^0\ra\Xi^-\pi^+)$ of $\sim (0.9-2)\%$~\cite{xic_predictions} result in ${\cal B}(B^+\ra\bar{\Xi}_c^0\Lambda_c^+)\sim (2.4-5.3)\times 10^{-3}$.
This can be compared with the theoretical prediction of $10^{-3}$~\cite{chernyak}.
On the other hand, the Belle measurement of ${\cal B}(\bar{B^0}\ra\Lambda_c^+\bar{p})=(2.19^{+0.56}_{-0.49}\pm 0.32\pm 0.57)\times 10^{-5}$~\cite{belle_lambdac_proton} is much smaller than 
their prediction of $4\times 10^{-4}$~\cite{chernyak}.
The very large ratio of $\sim 100$ for ${\cal B}(B\ra{\overline {\Xi}_c^0}\Lambda_c^+)/{\cal B}(\bar{B^0}\ra\Lambda_c^+\bar{p})$ disagrees with the naive expectation that the 
branching fractions for two-body baryonic $B$ decays proceeding via $b\ra c\bar{c}s$ and $b\ra c\bar{u}d$ transitions should be of the same order~\cite{chernyak}.

We thank the KEKB group for excellent operation of the
accelerator, the KEK cryogenics group for efficient solenoid
operations, and the KEK computer group and
the NII for valuable computing and Super-SINET network
support.  We acknowledge support from MEXT and JSPS (Japan);
ARC and DEST (Australia); NSFC (contract No.~10175071,
China); DST (India); the BK21 program of MOEHRD, and the
CHEP SRC and BR (grant No. R01-2005-000-10089-0) programs of
KOSEF (Korea); KBN (contract No.~2P03B 01324, Poland); MIST
(Russia); MHEST (Slovenia);  SNSF (Switzerland); NSC and MOE
(Taiwan); and DOE (USA).



\begin{thebibliography}{99}



\bibitem{belle_dpp}  Belle Collaboration, K.~Abe {\it et al}., Phys. Rev. Lett. {\bf 89}, 151802 (2002).

\bibitem{belle_kichimi_1} Belle Collaboration, N.~Gabyshev, H.~Kichimi {\it et al}., Phys. Rev. D {\bf 66}, 091102(R) (2002).

\bibitem{belle_ppk} Belle Collaboration, Y.-J.~Lee, M.-Z.~Wang {\it et al}., Phys. Rev. Lett. {\bf 93}, 211801 (2004);  M.-Z.~Wang, Y.-J.~Lee, et al.(Belle Collaboration), Phys. Rev. Lett. {\bf 90}, 201802 (2003); Belle Collaboration, K.~Abe {\it et al.}, Phys. Rev. Lett. {\bf 88}, 181803 (2002). 


\bibitem{kichimi_review}  H.~Kichimi, Nucl. Phys. B Proc. Suppl. {\bf 142}, 197 (2005).

\bibitem{theory_b_to_baryons} W.~S.~Hou and A.~Soni, Phys. Rev. Lett. {\bf 86}, 4247 (2001); C.~K.~Chua, W.~S.~Hou and S.~Y.~Tsai, Phys. Lett. B {\bf 528}, 233 (2002).

\bibitem{jpsilamp}  Belle Collaboration,  Q. L. Xie {\it et al}., Phys. Rev. D {\bf 72}, 051105(R) (2005).

\bibitem{kolya} Belle Collaboration, K.~Abe {\it et al}., hep-ex/0508015.


\bibitem{pdg} S.~Eidelman {\it et al}., Review of Particle Physics, Phys. Lett. B {\bf 592}, 1 (2004).


\bibitem{babar_incl_ws_lamc} BaBar Collaboration, B.Aubert {\it et al.}, Phys. Rev. D {\bf 70}, 091106 (2004).

\bibitem{dunietz}  I.~Dunietz, P.~Cooper, A.~Falk and M.~Wise, Phys. Rev. Lett. {\bf 73}, 1075 (1994).

\bibitem{chernyak} V.~L.~Chernyak and I.~R.~Zhitnitsky, Nucl. Phys B {\bf 345}, 137 (1990).

\bibitem{kekb} S.~Kurokawa and E.~Kikutani, Nucl. Instrum. Methods Phys. Res., Sect. A {\bf 499}, 1 (2003), and other papers included in this volume.

\bibitem{belle_detector} Belle Collaboration, A.~Abashian {\it et al}., Nucl. Instrum. Methods Phys. Res., Sect. A {\bf 479}, 117 (2002). 

\bibitem{svd2} Belle SVD2 Group, Y. Ushiroda Nucl. Instrum. Methods Phys. Res., Sect. A {\bf 511}, 6 (2003).

\bibitem{montecarlo} R.Brune {\it et al.}, GEANT 3.21, CERN DD/EE/84-1, 1984. 


\bibitem{id_pkpi}
Charged kaons are required to satisfy 
${\cal L}(K)/({\cal L}(K)+{\cal L}(\pi))>0.6$ and 
${\cal L}(K)/({\cal L}(K)+{\cal L}(p))>0.6$. 
Charged pions are required to satisfy 
${\cal L}(\pi)/({\cal L}(K)+{\cal L}(\pi))>0.1$ and 
${\cal L}(\pi)/({\cal L}(\pi)+{\cal L}(p))>0.1$. 
Protons are required to satisfy 
${\cal L}(p)/({\cal L}(K)+{\cal L}(p))>0.6$ and 
${\cal L}(p)/({\cal L}(\pi)+{\cal L}(p))>0.6$. 
Here ${\cal L}(K/\pi/p)$ is the 
particle identification likelihood for the $K/\pi/p$ hypotheses.
The above requirements have efficiencies of more than $95\%$ for pions, kaons and protons, respectively, from $B\ra\bar{\Xi}_c\Lambda_c^+$ decays. The probability for each particle species to be misidentified as one of the other two is less than $5\%$.  





\bibitem{fox2} G.~Fox and S.~Wolfram, Phys. Rev. Lett. {\bf 41}, 1581 (1978).


\bibitem{doublecounting} We found that after applying all the selection requirements, there are no multiple entries in the $M_{\rm bc}$ and $\Delta E$ distributions.


\bibitem{sidebands_definition} For the $\Xi_c^0$, the sidebands are determined as follows: 
$2.4<M(\bar{\Xi}_c^0)<2.44$~GeV/c$^2$ or $2.5<M(\bar{\Xi}_c^0)<2.54$~GeV/c$^2$.
For the $\Lambda_c^+$ the sidebands are determined as follows: 
$2.22<M(\Lambda_c^+)<2.26$~GeV/c$^2$ or $2.32<M(\Lambda_c^+)<2.36$~GeV/c$^2$. 


\bibitem{lesyak}   Belle Collaboration, T.~Lesiak {\it et al}., Phys. Lett. B {\bf 605}, 237 (2005), Erratum {\bf 617}, 198 (2005).

\bibitem{argus_function}  ARGUS Collaboration, H.~Albrecht {\it et al}., Phys. Lett. B {\bf 241}, 278 (1990).

\bibitem{xic_predictions} B.~Desplanques, J~.F.~Donoghue and B.~R.~Holstein, Annals Phys. {\bf 124}, 449 (1980); P.~Zenczykowski, Phys. Rev. D {\bf 40}, 2290 (1989); P.~Zenczykowski, Phys. Rev. D {\bf 50}, 402 (1994).

\bibitem{belle_lambdac_proton}   Belle Collaboration, N.~Gabyshev, H.~Kichimi, {\it et al}., Phys. Rev. Lett. {\bf 90}, 121802 (2003). 


\end{thebibliography}
\end{document}